\documentclass{article}
\usepackage{hiph-art}
\volnumber{19} \issuenumber{1} \edyear{2004}                             %|
\frompage{000} \topage{000}                                              %|
\recrevdate{1 January 2004}                                              %|
\def\rightmark{Dilepton Spectra at GSI-SIS Energies}
\def\leftmark{D. Schumacher, S. Vogel and M. Bleicher}

\newcommand{\inserttwoplots}[1]{\epsfysize=6.9cm\epsfbox{#1}}

\title{Theoretical Analysis of Dilepton Spectra\\
in Heavy Ion Collisions at GSI-FAIR Energies}
\authors{
{Diana Schumacher, Sascha Vogel and Marcus Bleicher %
\index{Schumacher, D.} % Abbreviated names of the author(s),
\index{Vogel, S.}
\index{Bleicher, M.} % to be inserted for use in the volume index
}\\[2.812mm]
{\normalsize
Institut f\"{u}r Theoretische Physik,\\
Johann Wolfgang Goethe-Universit\"at, Frankfurt am Main,\\
D-60438 Frankfurt am Main, Germany\\[0.2ex]
}}

\abstract{This paper addresses the theoretical analysis of dilepton spectra in C+C
collisions at GSI-SIS energies. Theoretical predictions for the recent data of the
HADES collaboration at SIS-energies are made with the help of a hadron-string transport
model, the Ultra-relativistic Quantum Molecular Dynamics (UrQMD) model. A mass shift of
the $\rho$ meson due to kinematical effects is discussed.}


\makeindex
\begin{document}

\maketitle

\section{Introduction}\label{intro}
Dilepton spectra are expected to play an important role in indicating deconfinement and
the restoration of the spontaneous breaking of chiral symmetry. Particularly in the low
mass region, dileptons couple directly to light vector mesons and can therefore reflect
the mesons mass distribution in the hot and dense region. Since dileptons do not
interact strongly with the surrounding hadronic medium, they escape the fireball of a
heavy ion collision and might carry information on properties of the early stage of the
collision. Thus, they are one of the most crucial observables for the exploration of
the QCD mass generation
\cite{Brown:1991kk,Rapp:1999ej,Cassing:1999es,Shekhter:2003xd,Cozma:2006vp}.

Heavy ion reactions with energies around $1-2~A$GeV create a reaction zone of high
nuclear density for about $10~$fm/c. This is long compared to the typical lifetime of a
hadronic resonance. Thus, the $\rho$-mesons with lifetimes of $\tau \simeq 1$~fm/c, can
decay within this zone and the decay products (i.e., the $e^{+} e^{-}$-pairs) still
carry the properties of the dense medium. The $\omega$ meson ($\tau \simeq 22$~fm/c)
and the $\phi$ meson ($\tau \simeq 44$~fm/c) have longer lifetimes and can therefore
escape the fireball before they decay. Consequently, the $\rho$ meson is the most
prominent candidate to study in-medium modifications of hadron properties in high
density nuclear collisions.

Calculations at low beam energies fall in the category of non-perturbative QCD, i.e.,
no exact methods to solve the equations of motion at QCD exist. Therefore one is forced
to utilise approximations within effective field theories and compare with the results
of experiments. In this paper, we focus on the HADES experiment at GSI-SIS
\cite{Schicker:1996nj}.

For our studies we apply the UrQMD model. It is a non-equilibrium transport approach
based on the covariant propagation of hadrons and strings. All cross sections are
calculated by the principle of detailed balance or are fitted to available data. The
model allows to study the full space time evolution for all hadrons, resonances and
their decay products. This permits to explore the emission patterns of the resonances
in detail and to gain insight into the origin of the resonances. UrQMD has been
successfully applied to study light and heavy ion reactions at SIS. Detailed
comparisons of UrQMD with a large body of experimental data at SIS energies can be
found in \cite{Sturm:2000dm}. For further details of the model the reader is referred
to \cite{Bass:1998ca,Bleicher:1999xi}.
\section{Theoretical Background for Dilepton Production}\label{section_theory}
\subsection{Dalitz Decay of $\pi^{0}$, $\eta$, $ \omega$ and $\eta'$}\label{subsec_dalitz}
Here, we follow \cite{Landsberg:1986fd} and \cite{Koch:1992sk} to derive the equations
for the decay rate of a meson into dileptons. In general, these processes have the form
\begin{equation} \label{Dalitz_decay3}
P \rightarrow V  e^{+} e^{-}, \quad V \rightarrow P e^{+} e^{-},
\end{equation}
with P being a pseudoscalar meson and V being a vector meson. There processes are not
true three-body decays because the processes can be split into two subprocesses. Thus,
the decay $A \rightarrow B e^{+} e^{-}$ can be separated into the decay of a virtual
photon, $A \rightarrow B \gamma^{\star}$, which subsequently decays via electromagnetic
conversion, $\gamma^{\star} \rightarrow e^{+} e^{-}$.  According to
\cite{Landsberg:1986fd} and \cite{Koch:1992sk} we arrive at the Dalitz decay formulas,
which differ only in the values for the form factors and if the particle B is a real
photon ($m_{B}=0$) or a virtual photon ($m_{B}\neq 0$):
\begin{itemize}
  \item Dalitz decay of the pseudoscalar mesons $\pi^{0}$, $\eta$ and
  $\eta'$ ($m_{B}=0$):\\
  \begin{eqnarray} \label{decay_rate_Dal_known1}
  \frac{dN_{A \rightarrow \gamma e^{+} e^{-}}}{dM} &=&
  \frac{4 \alpha}{3 \pi M} \sqrt{1 - \frac{4m_{e}^{2}}{M^{2}}} \left(1 + \frac{2
  m_{e}^{2}}{M^{2}} \right) \left(1 - \frac{M^{2}}{m_{A}^{2}} \right)^{3} \nonumber \\
  & & \times~~ |F_{AB}(M^{2})|^{2} \frac{\Gamma_{A \rightarrow 2 \gamma}}{\Gamma_{tot}} \langle N_{A} \rangle.
  \end{eqnarray}
  The quantity $\langle N_{A} \rangle$ is the number of mesons A per event, M is the invariant mass of the lepton pair, $m_e$ the electron mass, and $F_{AB}$ the corresponding form factor.
  \item Dalitz decay of the vector meson $\omega$ ($m_{B} \neq 0$):\\
  \begin{eqnarray} \label{decay_rate_Dal_known2}
  \frac{dN_{A \rightarrow B e^{+} e^{-}}}{dM}&=&\frac{2 \alpha}{3 \pi M}
  \sqrt{1 - \frac{4m_{e}^{2}}{M^{2}}} \left(1 + \frac{2m_{e}^{2}}{M^{2}} \right)
  |F_{AB}(M^{2})|^{2} \frac{\Gamma_{A \rightarrow 2 \gamma}}{\Gamma_{tot}}
  \langle N_{A} \rangle \nonumber \\
  & & \times \left( \left(1 + \frac{M^{2}}{m_{A}^{2}-m_{B}^{2}}\right)^{2}
  - \left(\frac{2m_{A}M}{m_{A}^{2}-m_{B}^{2}}\right)^{2} \right)^{3/2}.
  \end{eqnarray}
\end{itemize}

The Dalitz decay of the $\Delta$ resonance, $\Delta \rightarrow N e^{+} e^{-}$, differs
from the previous discussed decays because of the different interaction $N \Delta
\gamma$ vertex. The decay rate reads as follows
\begin{equation} \label{dNdM rate_Delta}
\frac{dN_{e^{+} e^{-}}}{dM} = \int \frac{dN_{\Delta \rightarrow N e^{+}
e^{-}}}{dM}(M_{\Delta}) \frac{dN_{\Delta}}{d M_{\Delta}} d M_{\Delta} = \int
\frac{2\alpha}{3 \pi M} \frac{\Gamma(M_{\Delta}, M)}{\Gamma_{\Delta 0}^{tot}}
\frac{dN_{\Delta}}{d M_{\Delta}} dM_{\Delta},
\end{equation}
where the fraction $\frac{dN_{\Delta \rightarrow N e^{+}
e^{-}}}{dM}(M_{\Delta})=\frac{d\Gamma_{\Delta \rightarrow N e^{+} e^{-}}}{dM
\quad\Gamma_{\Delta 0}^{tot}}(M_{\Delta})$ is the number of $e^{+} e^{-}$-pairs per
$\Delta$ resonance, invariant mass bin and event. $M_{\Delta}$ is the actual mass of
the $\Delta$ resonance from the simulation and $\Gamma_{\Delta 0}^{tot}$ is the total
decay width at the resonance pole mass. Additionally, we use the decay width into a
massive (virtual) photon
\begin{equation} \label{decay_Delta1}
\Gamma(M_{\Delta}, M) = \frac{\lambda^{1/2}(M^{2}, m_{N}^{2}, M_{\Delta}^{2})}{16 \pi
M_{\Delta}^{2}} m_{N} [2 \mathcal{M}_{t}(M, M_{\Delta}) + \mathcal{M}_{l}(M,
M_{\Delta})].
\end{equation}
The function $\lambda$ is defined by $\lambda(m_{A}^{2}, m_{1}^{2}, m_{2}^{2})
 = (m_{A}^{2}-(m_{1}+m_{2})^{2})(m_{A}^{2}-(m_{1}-m_{2})^{2})$. The corresponding matrix
elements $\mathcal{M}_{t}$ and $\mathcal{M}_{l}$ are taken from \cite{Wolf:1990ur}.
\subsection{Direct Decay}\label{subsec_direct}
The decay of a neutral vector meson in an $e^{+} e^{-}$-pair is a true two-body decay.
Thus, energy and momentum conservation leads to the fact that the mass of the original
meson (and the virtual photon) is the same as the invariant mass of the dilepton
($m_{V} = M_{e^{+} e^{-}}$). For a direct decay the branching ratio reads
\cite{Koch:1992sk}
%\begin{equation} \label{BR_direct}
$BR (V \rightarrow e^{+} e^{-}) = \frac{\Gamma_{V \rightarrow e^{+}
e^{-}}(M)}{\Gamma_{tot}}$
%\end{equation}
with $V = \rho^{0},~ \omega,~ \phi$. $\Gamma_{V \rightarrow e^{+} e^{-}}$ varies with
the dilepton mass like $M^{-3}$ according to \cite{Ko:1996is}
\begin{equation}
\Gamma_{V \rightarrow e^{+} e^{-}}(M) = \frac{\Gamma_{V \rightarrow e^{+}
e^{-}}(m_{V})}{m_{V}} \frac{m_{V}^{4}}{M^{3}} \sqrt{1 - \frac{4m_{e}^{2}}{M^{2}}}
\left(1 + 2 \frac{m_{e}^{2}}{M^{2}} \right)
%\nonumber\\
%& \simeq & \frac{\Gamma_{V \rightarrow e^{+} e^{-}}(m_{V})}{m_{V}}
%\frac{m_{V}^{4}}{M^{3}}
\end{equation}
with $\Gamma_{V \rightarrow e^{+} e^{-}}(m_{V})$ the partial decay width at the
resonance peak mass. For each meson mass the branching ratio ($BR$) is calculated and
summed over in each mass bin $\Delta M$ to get the distributions $dN/dM$ for the
figures shown below.

The present calculations do not consider contributions from bremsstrahlung, because of
the relatively small rates \cite{Ernst:1997yy}.
\subsection{Implementation into the Simulation}\label{subsec_implementation}
In order to extract the dilepton abundances from the UrQMD model
\cite{Bass:1998ca,Bleicher:1999xi}, we use the lifetimes and 4-momenta of the
calculation for each channel. A meson can be produced via the decay of a meson or
baryon resonance, annihilation or string fragmentation. Such a meson obtains its mass
according to the relativistic Breit-Wigner distribution, and a 4-momentum vector
resulting from the kinematics in the single process under consideration. After the
creation, the meson is assumed to travel on a straight line until it decays or it
collides with another particle and its 4-momentum changes.

The most time consuming factor with the numerical analysis of electromagnetic processes
is the suppression by a factor of $\alpha^{2} \approx 10^{-5}$ due to the
electromagnetic coupling. Therefore, we assume that the vector mesons radiate dileptons
continuously during their whole lifetime (``shining'' method). After the global
freeze-out time $t_{f}$, the dilepton yield can be treated as in \cite{Wolf:1990ur} and
\cite{Heinz:1991fn} and we get
\begin{equation} \label{time_dil_yield}
\frac{dN_{e^{+} e^{-}}}{dM} = \int_{0}^{t_{f}} dt~ \frac{dN_{V}(t)}{dM} \Gamma_{V
\rightarrow e^{+} e^{-}}(M) + \frac{\Gamma_{e^{+} e^{-}}(M)}{\Gamma_{tot}^{V}(t_{f})}
\frac{dN_{V}(t_{f})}{dM}.
\end{equation}
The second term of Eq. (\ref{time_dil_yield}) describes the process very well if the
decay rates do not depend on time. The invariant mass distribution of the dileptons is
then equal to the distribution of the meson (V) itself multiplied with the branching
ratio. But in a time dependent simulation one has to consider collisional broadening,
i.e., the fact that the dilepton can be reabsorbed before it decays. This means that
the lifetime $\tau$ of the vector meson producing the lepton pair and the decay width
$\Gamma$ varies for each considered hadron. In this case it is more suitable to use the
time integration method.
\section{Dilepton Yields at SIS Energies}\label{section_HADES}
In this section we present predictions for dilepton spectra which correspond to the
experimental runs carried out by the HADES collaboration \cite{HADES}. We consider
minimum bias (b $\leq$ 4.5~fm) C+C collisions at $2~A$GeV. For the calculation we
utilise the UrQMD model. To achieve better statistics, we randomly create 1 to 100
dilepton pairs per hadron decay (depending on hadron species and original abundances)
with the direction and momenta of the dileptons in the local rest frame of each hadron.
These $e^{+} e^{-}$ momenta are transformed to the laboratory frame. For convenience,
all spectra
are then normalised to the mean pion multiplicity $\langle N_{\pi^{0}} \rangle$. \\

\begin{figure}[hbt]
%
%\vspace*{-0.8cm}
\inserttwoplots{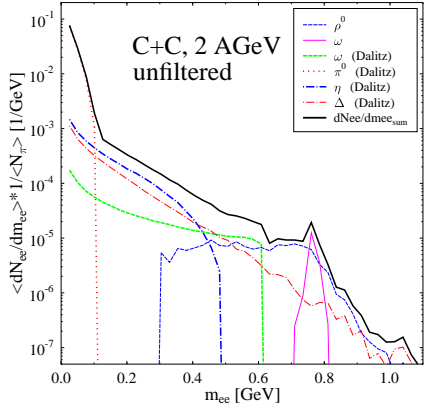}
%\vspace*{-1.5cm}
%
%\caption[]{\label{mass_spectra_2gev} Invariant mass spectra for dilepton production in
%the reaction C+C at $2~A$GeV.}
%
\hfill
%
%\vspace*{-1.6cm}
\inserttwoplots{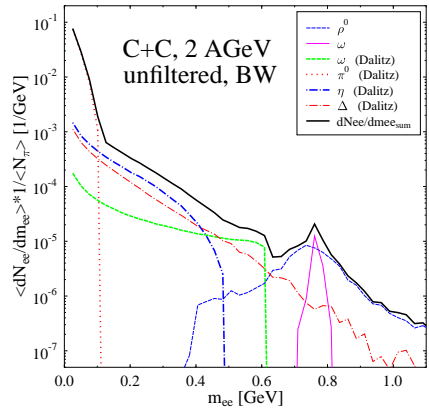}
%\vspace*{-1.5cm}
%
\caption[]{Invariant mass spectra for dilepton abundances in minimum bias C+C
collisions at $2~A$GeV (left). The yields are normalised to the pion multiplicity
$\langle N_{\pi^{0}} \rangle$ = 1.223/event. In the right panel the $\rho$ meson is
distributed via a free Breit-Wigner distribution.} \label{mass_spectra_2gev}
\end{figure}

The left hand side of Fig. \ref{mass_spectra_2gev} depicts the invariant dilepton mass
spectra of C+C collisions at $2~A$GeV. The pion multiplicity obtained from the UrQMD
calculation is $\langle N_{\pi^{0}} \rangle$ = 1.223/event. One observes that for
masses below 110~MeV the $\pi^{0}$-Dalitz decay contribution is up to two magnitudes
higher than the other contributions but does not account for higher masses. The
$\Delta$- and $\eta$-Dalitz decay outweigh the other yields until 400~MeV. For masses
from 400 to 600~MeV the $\omega$-Dalitz and direct $\rho$ decay contributions become
important. The direct decays of the $\rho$ and also that of the $\omega$ are
responsible for the yields with the highest masses up to 1~GeV.\\
We include the coupling of the $\rho$ meson to pions and baryons via the employed cross
sections fitted to available data or calculated from detailed balance
\cite{Bass:1998ca,Bleicher:1999xi}. In the intermediate-mass region around 500 to 600
MeV the $\rho$ spectrum is enhanced due to the decay chain $N^{*}(1520) \rightarrow N +
\rho;~~ \rho \rightarrow e^{+} e^{-}$. For higher $\rho$ masses heavier baryon
resonance decays and pion-pion annihilation are the most contributing parts to the
spectrum. This leads to a small second hump in the dilepton spectrum. One hump is
centered around 500 to 600~MeV and one at the $\rho$ meson vacuum mass
($m_{\rho}=770~$MeV).

Although the present calculation is based on vacuum cross sections and vacuum hadron
widths the spectra from this many body dynamics look like what is expected from
calculations employing in-medium spectral functions explicitly \cite{Cozma:2006vp}. Due
to the decay kinematics of the production channel of the $\rho$ meson a strong
modification of the $\rho$ spectral function occurs. This effect is of kinematic origin
only and will be discussed in the following section. \\

To investigate the difference between the $\rho$ meson distribution originating from
the UrQMD simulation with all the kinematics applied and the vacuum one, we compare now
the spectra above to the case that the $\rho$ mass has a free Breit-Wigner
distribution. Figure \ref{mass_spectra_2gev} (right) depicts this scenario, the other
rates remain unaltered. The discrepancy is obvious: The low-mass part of the $\rho$ is
suppressed whereas masses over 900~MeV appear. Additionally, a clear maximum around the
$\rho$ vacuum mass of 770~MeV is visible. This leads to a dip in the sum of all rates
around 600~MeV. The original UrQMD spectra, in contrast, show no maximum at the vacuum
mass. It is an approximately flat curve for $2~A$GeV. In conclusion, the present
transport model simulations suggest strong kinematic effects and leads in fact to
modifications of the $\rho$ mass spectrum and dilepton spectra. \\
\begin{figure}[hbt]
\vspace*{-0.8cm} \insertplot{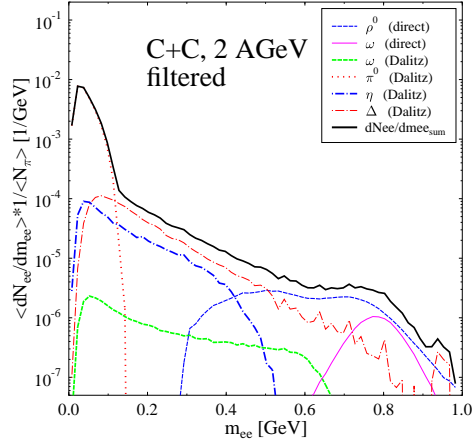} \vspace*{-1.5cm}
%
%\caption[]{\label{mass_spectra_2gev} Invariant mass spectra for dilepton production in
%the reaction C+C at $2~A$GeV.}
%
\hfill
%
%\vspace*{-1.6cm}
%\inserttwoplots{dNeedmee_expHADES.eps}
%\vspace*{-1.5cm}
%
\caption[]{Invariant mass spectrum for dilepton production in minimum bias C+C collisions
at $2~A$GeV. The yields are filtered by the acceptance matrix of the HADES
collaboration, release 1.0.} \label{mass_spectra_filter}
\end{figure}
Fig. \ref{mass_spectra_filter} depicts invariant dilepton mass spectra from C+C
collisions at $2~A$GeV corrected to the HADES acceptance. The calculated dilepton
momentum vectors were filtered by the HADES collaboration (the official version of the
HADES acceptance filter, release 1.0, was used). This filter contains an acceptance
matrix for each point at full solid angle and a smearing of the momenta according to
experimental resolution. \\
\section{$\rho$ meson contributions}

\begin{figure}[hbt]
\vspace*{-0.8cm} \insertplot{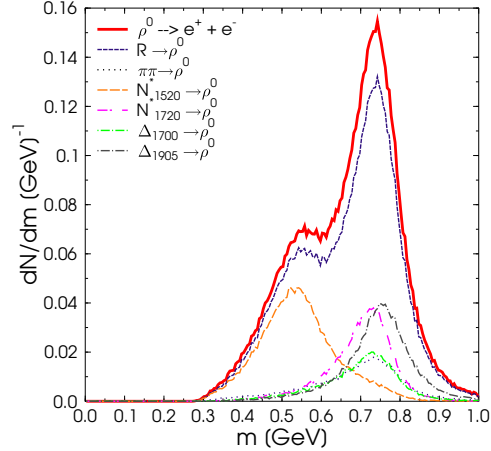} \vspace*{-1.5cm}
%
%\caption[]{\label{mass_spectra_2gev} Invariant mass spectra for dilepton production in
%the reaction C+C at $2~A$GeV.}
%
\hfill
%
%\vspace*{-1.6cm}
%\vspace*{-1.5cm}
%
\caption[]{Mass distribution for $\rho^0$ mesons for minimum bias C+C reactions at 2~\textit{A}GeV.
The peak around 500~MeV is due to a strong contribution from
$N^*_{1520} \rightarrow p+\rho$ which amounts to 75\% for masses below 600~MeV. The full line depicts the mass distribution of all decayed $\rho$ mesons. The dashed line right below the full line depicts the $\rho$ mesons originating from resonance decays, whereas the dotted line depicts the $\rho$ mesons originating from $\pi\pi$ annihilation. The dashed line centered around $\sim$ 500~MeV depicts the $\rho$ mesons originating from $N^*(1520)$ decays.} \label{mass_detailed}
\end{figure}

Let us finally discuss the $\rho$ meson contributions to the dilepton spectrum. Fig.
\ref{mass_detailed} depicts the $\rho$ meson mass spectrum for minimum bias C+C
reactions at 2~\textit{A}GeV beam energy. One observes a double peak structure, where
the peak at $\sim$ 500~MeV originates from $N^*(1520)$ resonance decays, i.e., by the
decay chain $N^*(1520) \rightarrow N + \rho$. As seen in Fig. \ref{mass_detailed}
$\rho$ mesons originating from such a baryon resonance decay have masses below 600~MeV.
This effect is of kinematic nature only, since the calculation does not make use of
explicit in-medium spectral functions. $\rho$ mesons originating from $\pi\pi$
annihilation or higher mass baryon resonance decays have a mass, which is compatible
with the pole mass of 770~MeV. Note that the $\rho$ production is dominated by the
decay of baryonic resonances. Direct $\rho$ production from $\pi \pi$ scattering yields
only a small contribution up to 20 \% at the $\rho$-peak mass. For more details
regarding $\rho$ mesons in C+C collisions at 2~\textit{A}GeV beam energy, the reader is
referred to \cite{Vogel:2005pd}.

\section{Conclusions}\label{concl}
We have shown that due to the collision kinematics the mass spectrum of vector mesons,
especially of the $\rho$, is modified. Therefore, a many body transport approach
without explicit use of in-medium spectral functions leads to sizeable shifts in the
dilepton mass distribution compared to a vacuum baseline. The prediction of a
kinematically induced mass shift at 2~\textit{A}GeV beam energy can be tested by the
HADES experiment, e.g. when going towards target/projectile rapidities.
\section*{Acknowledgments}
This work has been supported by BMBF, DFG and GSI. Fruitful discussions with Christian
Sturm, Joachim Stroth and Horst St\"ocker are gratefully acknowledged. Computational
resources have been provided by the Center of Scientific Computing (CSC) at Frankfurt
am Main.

\vfill\eject
\end{document}